\documentclass[a4paper,fleqn]{cas-dc}

\usepackage{ragged2e}  

\usepackage[numbers]{natbib}
\usepackage{verbatim, amsmath, amsfonts, amssymb, amsthm}
\usepackage{graphicx}
\usepackage{epstopdf}
\usepackage{acronym}
\usepackage{booktabs}
\usepackage{moreverb,url}
\usepackage{placeins}
\usepackage{soul}
\usepackage{siunitx}
\usepackage{acronym}
\usepackage{float}
\usepackage{hhline}
\usepackage{pdflscape}   
\usepackage{array}       
\usepackage{tabularx}
\usepackage{booktabs}
\usepackage[margin=6pt]{subfig}
\usepackage{balance}
\usepackage{enumitem}
\usepackage{makecell}

\newcommand\BibTeX{{\rmfamily B\kern-.05em \textsc{i\kern-.025em b}\kern-.08em
T\kern-.1667em\lower.7ex\hbox{E}\kern-.125emX}}
\usepackage{color}
\usepackage{tikz}
\usepackage{adjustbox}

\usepackage{xurl}     

\acrodef{ADA}{American diabetes association}
\acrodef{EASD}{European association for the study of diabetes}
\acrodef{MPC}{model predictive control}
\acrodef{EKF}{extended Kalman filter}
\acrodef{RMSE}{root mean square error}
\acrodef{IOB}{insulin on board}
\acrodef{BG}{blood glucose}
\acrodef{APS}{artificial pancreas systems}
\acrodef{AID}{automated insulin delivery}
\acrodef{AP}{Artificial Pancreas}
\acrodef{CLC}{closed-loop control}
\acrodef{T1D}{type 1 diabetes}
\acrodef{HbA1c}{hemoglobin A1c}
\acrodef{TDI}{total daily insulin}
\acrodef{IIT}{intensive insulin therapy}
\acrodef{CR}{carbohydrate-to-insulin ratio}
\acrodef{CF}{correction factor}
\acrodef{CGM}{continuous glucose monitor}
\acrodef{TIR}{time in range}
\acrodef{TITR}{time in tight range}
\acrodef{TAR}{time above range}
\acrodef{TBR}{time below range}
\acrodef{PID}{proportional-integral-derivative}
\acrodef{CVGA}{control variability grid analysis}
\acrodef{FDA}{food and drug administration}
\acrodef{LBGI}{low blood glucose index}
\acrodef{HBGI}{high blood glucose index}
\acrodef{DIA}{duration of insulin action}
\acrodef{SMBG}{self-monitoring blood glucose}
\acrodef{CHO}{carbohydrate}
\acrodef{IIR}{insulin infusion rate}
\acrodef{PK}{pharmacokinetic}
\acrodef{EGP}{endogenous glucose production}
\acrodef{IG}{interstitial glucose}
\acrodef{SD}{standard deviation}
\acrodef{PDF}{probability density function}
\acrodef{GIR}{glucose infusion rate}
\acrodef{MSE}{mean squared error}
\acrodef{NE}{net effect}
\acrodef{IVS}{insulin variability signal}
\acrodef{IBD}{insulin bolus delivery}
\acrodef{LTI}{linear time invariant}
\acrodef{DSS}{decision support systems}
\acrodef{MDI}{multiple daily injections}
\acrodef{MAS}{meal-autonomous system}
\acrodef{hAID}{hybrid AID}
\acrodef{RoC}{rate of change}
\acrodef{UniBE}{University of Bern}

\def\tsc#1{\csdef{#1}{\textsc{\lowercase{#1}}\xspace}}
\tsc{WGM}
\tsc{QE}
\tsc{EP}
\tsc{PMS}
\tsc{BEC}
\tsc{DE}

\newcommand{\umin}{u_k^{\min}}
\newcommand{\umax}{u_k^{\max}}
\newcommand{\deltaumin}{\Delta u^{\min}}
\newcommand{\deltaumax}{\Delta u^{\max}}
\begin{document}
\let\WriteBookmarks\relax
\def\floatpagepagefraction{1}
\def\textpagefraction{.001}

\shorttitle{Advanced Hybrid Artificial Pancreas}
\shortauthors{V Naik, E. Manzoni et~al.}

\title [mode = title]{Advanced Hybrid Automated Insulin Delivery System based on Successive Linearization Model Predictive Control: The UniBE System}                   

\tnotetext[1]{V. Naik and E. Manzoni should be considered joint first authors. This work was supported by internal seed grant funds from the University of Bern awarded to Jose Garcia-Tirado.}

\author[1,2]{Vihangkumar V. Naik}[orcid=0000-0002-2214-8052]
\ead{vihangkumar.naik@unibe.ch}

\author[1,2]{Eleonora Manzoni}[orcid=0000-0002-2445-6104]
\ead{eleonora.manzoni@unibe.ch}

\author[1,2,3]{Clara Escorihuela-Altaba}[orcid=0009-0003-1680-920X]
\ead{clara.escorihuelaaltaba@students.unibe.ch}

\author[1,2]{Jose Garcia-Tirado}[orcid=0000-0002-9970-2162]
\cormark[1]
\ead{jose.garcia@unibe.ch}

\address[1]{Department of Diabetes, Endocrinology, Nutritional Medicine, and Metabolism, Inselspital, Bern University Hospital and University of Bern, Switzerland.}

\address[2]{ Diabetes Center Berne, Bern, Switzerland.}

\address[3]{ Graduate School for Cellular and Biomedical Sciences, University of Bern, Bern, Switzerland.}

\cortext[cor1]{Corresponding author}

\begin{abstract}
\noindent \textit{Background and objective:} 
Hybrid automated insulin delivery (hAID) systems represent the most advanced therapy for type 1 diabetes (T1D). Current systems rely on linear or linearized models of glucose homeostasis, which may compromise prediction accuracy and, in turn, timely decision-making by the controller. Physiological variability further complicates insulin requirements, underscoring the need for controllers that adapt dynamically and reduce user burden. 

\noindent \textit{Methods:} 
We introduce the \ac{UniBE} hAID system, a framework based on successive linearization \ac{MPC}. The controller integrates basal insulin infusion with the insulin bolus delivery module for meal-related and corrective bolus dosing, adapting bounds in real time to glucose dynamics while accounting for both automated and user-initiated inputs. In-silico evaluation was conducted using the commercial version of the FDA-accepted UVa/Padova metabolic simulator across nine scenarios involving persistent and time-varying errors in meal timing, carbohydrate estimation, and basal insulin profiles. 

\noindent \textit{Results:} 
In the baseline scenario, \ac{UniBE} achieved a mean time in range of 92.0$\pm$13.2\%, with time below range at 0.1$\pm$0.2\% and time above range at 7.9$\pm$13.2\%. Across perturbation scenarios, time in range remained between 75.1–92.8\%, with low hypoglycemia incidence, demonstrating resilience to clinically relevant disturbances. 

\noindent \textit{Conclusion:} 
The \ac{UniBE} hAID system maintained stable glycemic regulation across diverse error conditions, highlighting its robustness and adaptability. These findings establish proof-of-concept feasibility and provide a foundation for further refinement, although these do not constitute a full pre-clinical validation. Future work must extend beyond in silico testing to include more preclinical and clinical studies to rigorously assess safety, efficacy, and real-world performance. 
\end{abstract}

\begin{highlights}
\item The UniBE hybrid automated insulin delivery (hAID) system using successive linearization model predictive control. It integrates basal insulin modulation with meal-related and hyperglycemia corrective bolus dosing for effective insulin management.

\item The controller glycemic regulation performance is evaluated in nine clinically relevant perturbation scenarios with the commercial version of the FDA-accepted UVa/Padova simulator.

\item The UniBE hAID achieved high time in range (on average 75-93\%) with minimal hypoglycemia across diverse challenging scenarios.

\item This work demonstrates feasibility, establishing a foundation for preclinical and clinical validation.
\end{highlights}


\maketitle
\sloppy
\section{Introduction}
\label{sec:introduction}

\Ac{T1D} is a chronic autoimmune disease requiring lifelong insulin therapy and remains a global health concern \cite{gregory2022global}. Effective management of \ac{T1D} requires continuous glucose monitoring and precise insulin dosing, yet conventional therapy places a heavy burden on patients and caregivers \cite{ADA2025Technology}. Despite these efforts, hypoglycemia and hyperglycemia remain common, driving both acute risks and long‑term complications~\cite{AdverseOutcomesT1D2024}. These challenges underscore the need for therapeutic strategies that ease patient workload while ensuring safe, reliable glycemic control \cite{sherr2022automated}.

\Ac{AID} systems are increasingly recognized by the \ac{ADA} and \ac{EASD} as the emerging standard of care for \ac{T1D} \cite{ADA2025Technology, ADA_EASD2026}, offering improved \ac{BG} regulation and quality of life \cite{Breton2021ControlIQ,DaSilva2022MiniMed780G,Forlenza2023Omnipod5}. Operating in closed-loop, \ac{AID} systems calculate insulin requirements from \ac{CGM} data and deliver doses via a subcutaneous insulin pump \cite{aiello2021review}. Modern \ac{CGM} sensors provide glucose readings every 1–5 minutes for up to 10-15 days \cite{lin2021continuous}, while the control unit adjusts infusion rates using real-time sensor data and often patient-reported information \cite{bequette2014fault}.

Within the spectrum of \ac{AID} systems, a distinction can be made between \ac{MAS}, which aim to fully automate insulin dosing, including managing meal-related glucose excursions, and hybrid systems, which automate basal insulin delivery and corrective boluses but still require user-initiated meal boluses \cite{ware2022closed}. Meal-autonomous designs remain experimental due to the difficulty of reliably detecting and compensating for meals, whereas \ac{hAID} systems are clinically established \cite{castle2017future,boughton2024role}. When compared to multiple daily injections (MDI) or sensor-augmented pump (SAP), \ac{hAID} offers clear advantages, e.g., improved time in safe euglycemic range, reduced hypoglycemia, and enhanced safety through user input at critical points. Clinical trials of popular \ac{hAID} technologies, such as Control‑IQ, MiniMed 780G, and Omnipod 5, consistently confirm significant gains in glycemic outcomes and quality of life, underscoring their role as the current standard of care for \ac{T1D} \cite{brown2019six,silva2022real,leelarathna2025efficacy}.

Control algorithms are central to \ac{hAID} performance \cite{thomas2022algorithms,cinar2019automated}. Early systems employed \ac{PID} controllers, which adjust insulin delivery based on glucose deviation, accumulated error, and rate of change \cite{steil2008closed,steil2013algorithms}. While simple and efficient, \ac{PID} controllers are vulnerable to sensor noise and delays \cite{Pinster2016MPCvsPID}. Fuzzy logic has also been explored, applying rule-based reasoning to physiological uncertainty \cite{nimri2014night}. More recently, \ac{MPC} has gained prominence, using mathematical models of glucose–insulin dynamics to predict future states and optimize dosing under safety constraints. \ac{MPC} is particularly effective in managing variability and delays and underpins several advanced systems \cite{Breton2012AP,Hovorka2004NMPC,Turksoy2018MultimoduleAP,Pinster2016MPCvsPID,garcia2021advanced}. However, all available designs rely on linear or linearized models of glucose homeostasis, thereby compromising prediction accuracy in favor of computational feasibility.

In this work, we introduce the University of Bern (UniBE) \ac{hAID} system, which employs a successive-linearization \ac{MPC} framework for automated insulin delivery. This approach extends conventional \ac{MPC} by repeatedly linearizing the nonlinear glucose–insulin dynamics around the current operating point, yielding a linear \ac{MPC} problem that is solved at each sampling instance~\cite{Henson1998NMPCReview}. This enables the controller to adapt to changing dynamic conditions and inherent nonlinearities in real-time, leveraging computation efficiency as each step involves solving a standard quadratic programming problem.
Furthermore, the proposed \ac{MPC} formulation enables systematic integration of actionable patient information to adapt the controller behavior for personalized insulin therapy. In particular, the controller incorporates adaptive constraints on the optimal insulin dosage that account for dynamical changes in glucose levels, bolus dosing, and patient-specific information.  
The dynamical model employed in the controller is personalized by identifying a reduced set of the most impactful model parameters. This is achieved by following the methodology proposed by the authors in~\cite{Escorihuela2025ParameterRelevance}, which provides a systematic means of ranking parameters according to their influence on model output using an extended Sobol sensitivity analysis~\cite{Saltelli2008SensitivityPrimer}.

The \ac{UniBE} \ac{AID} system pairs a core \ac{MPC} controller with successive linearization for basal insulin modulation with an \ac{IBD} module for prandial and corrective boluses. Its architecture emphasizes robustness, aiming to sustain glycemic control under diverse and imperfect conditions that reflect everyday challenges. To evaluate the system's performance, we used the commercial version of the \ac{FDA}‑accepted UVa/Padova metabolic simulator in our in‑silico experimentation \cite{Man2014UVAPadovaSimulator}. Within this framework, we introduced perturbations such as errors in basal insulin estimation, inaccuracies in \ac{CHO}  counting and communication time, and deviations in initial glycemic states. These scenarios were chosen to approximate common sources of error in diabetes management, thereby testing the resilience of the \ac{UniBE} controller under conditions that mimic, but do not fully replicate, real-world variability. 

This work should be regarded as a proof-of-concept rather than a definitive pre-clinical validation. The \ac{UniBE} controller was intentionally designed to address challenging scenarios, and the in‑silico experiments demonstrate its ability to adapt successfully across diverse perturbations. By systematically introducing variability, we aimed to evaluate the algorithm's robustness under conditions that reflect everyday \ac{T1D} experience.

The contribution of this study lies in demonstrating closed-loop feasibility: the controller can sustain performance across scenarios that emulate common errors and variability in diabetes management. These findings establish a foundation for further refinement and motivate subsequent steps toward experimental validation. However, they should not be interpreted as conclusive evidence of clinical efficacy. Future work must extend beyond in‑silico testing to include preclinical and clinical studies, where the true adaptability and safety of the controller can be rigorously assessed.

\section{Methods}
\label{sec:Methods}
\subsection{The UniBE control system }
The \ac{UniBE} control system is an advanced \ac{hAID}  system that integrates \ac{CGM}, an interoperable insulin pump, a successive linearization \ac{MPC} algorithm, and an IBD to maintain glucose levels within near-euglycemic ranges. Below, we describe the main ingredients of the \ac{UniBE} controller architecture.

\subsubsection{The glucose regulatory model}
The model considered in the \ac{MPC} architecture is the Hovorka glucose-insulin model \cite{Hovorka2002IVGTTPartitioning, Hovorka2004NMPC} which consists of the subcutaneous insulin absorption submodel, the glucose-regulatory system submodel, and the \ac{CHO} absorption submodel \cite{Wilinska2010Simulation}. Details about this mathematical model can be found in the Appendix. 

The dynamical model is personalized by narrowing down a set of the most impactful parameters and then finding the physiologically plausible values that allow the description of the user’s \ac{CGM} values via methods from system identification. The parameter selection for personalization follows the methodology proposed by the authors in~\cite{Escorihuela2025ParameterRelevance}, ranking the parameters based on their influence on model output using an extended Sobol sensitivity analysis \cite{Saltelli2008SensitivityPrimer}.

\subsubsection{The \ac{UniBE} \ac{MPC} unit}
\label{subsec: The UniBE Model Predictive Control (MPC) System }
The mathematical formulation of the \ac{UniBE} \ac{MPC} unit is summarized as follows: 
\begin{equation} \label{eq:MPCoptimizationObjective}
    \min_{U} \sum_{k=0}^{N_p - 1} (y_k - r_k)^T Q_{\text{mpc}} (y_k - r_k) + \sum_{k=0}^{N_c - 1} \Delta u_k^T R_{\text{mpc}} \Delta u_k
\end{equation}
s.t.:
\begin{equation} \label{eq:systemDynamics1}
x_{k+1} = A x_k + B u_k, \quad k = 0, \ldots, N_p - 1
\end{equation}
\begin{equation} \label{eq:systemDynamics2}
y_k = C x_k, \quad k = 0, \ldots, N_p - 1
\end{equation}
\begin{equation} \label{eq:constraint1}
\umin \leq u_k \leq \umax, \quad k = 0, \ldots, N_c - 1
\end{equation}
\begin{equation} \label{eq:constraint2}
\resizebox{\columnwidth}{!}{$
[\umin, \umax] =
\begin{cases}
[0, N_{\text{TDI}}^{\max} \text{TDI}], & \text{if } \text{IoB} \leq N_{\text{TDI}}^{\max}\text{TDI} \\
[u_{\text{basal}}, N_{\text{TDI}}^{\min} \text{TDI}], & \text{otherwise}
\end{cases}
$}
\end{equation}

where \eqref{eq:MPCoptimizationObjective} is the \ac{MPC} optimization objective subject to system dynamics \eqref{eq:systemDynamics1} \eqref{eq:systemDynamics2}, and constraints \eqref{eq:constraint1} \eqref{eq:constraint2}, with $U=[u_0, \dots u_{N_c-1} ]$ a vector of future insulin injections; $\umin, \umax$ the lower and upper bounds to the insulin vector; $\deltaumin$ and $\deltaumax$ the bounds for $\Delta u_k = u_k - u_{k-1}$; $N_p$ and $N_c$ are the prediction and control horizons, respectively; $Q_{\text{mpc}}=\tfrac{Q}{u^2_{\text{basal}}}$ and $R_{\text{mpc}}=\tfrac{R}{u^2_{\text{basal}}}$  weight the costs associated with the glucose target and regularization term, respectively, and are functions of the user’s basal insulin rate ($u_{\text{basal}}$); $y_k$ the predicted glucose; $r_k$ is the glucose target. 

Constraints \eqref{eq:systemDynamics1} and \eqref{eq:systemDynamics2} enforce the personalized model of the user, with the duplet $(A_k,B_k)$ of personalized, discrete-time, state and input matrices updated at each time instance after successive linearization \cite{Henson1998NMPCReview} around the current state provided by a personalized \ac{EKF} \cite{Welch1995KalmanFilter}. For simplicity, the subscript is dropped in \eqref{eq:systemDynamics1}. Constraint \eqref{eq:constraint1}  ensures the physical and operational feasibility of the insulin injections within the optimization problem. Equation \eqref{eq:constraint2} defines the dynamic nature of the insulin bounds as a function of the \ac{TDI}  for basal insulin, i.e., the total amount of basal insulin required over a 24-hour period, and the scaling hyperparameters $N_{\text{TDI}}^{\min}$ and $N_{\text{TDI}}^{\max}$. It is important to note that such hyperparameters should be selected with care, following clinical guidelines to ensure safety, and should never be set to excessively high values. In this equation, IoB represents the insulin on board as the estimated amount of active insulin remaining from previous insulin injections. 
This provide a systemic means of adapting degree of freedom of the controller actions $u_k$ by accounting for the estimates of insulin already active in the body administered by manual-auto correction as well as prandial boluses.

In addition to Equations \eqref{eq:constraint1}, \eqref{eq:constraint2}, the controller bounds are also adapted as a function of the sensor-measured BG levels and their \ac{RoC} as follows:
\begin{equation} \label{eq:bounds}
\begin{array}{@{}l@{}}
[\umin, \umax] = \\[1ex]
\begin{cases}
[0, u_{\text{basal}}], & \text{if}
  \begin{array}[t]{l}
    y_k^{\mathrm{cgm}} \in (70, 120]
  \end{array} \\
[0, 0], & \text{if}
  \begin{array}[t]{l}
    y_k^{\text{cgm}} \in (\text{low}, 70] \\
 \cup \bigl\{ y_k^{\text{cgm}} \in (70, 200) \mid \\
 \mathrm{RoC}_k \leq -N_{\mathrm{RoC}}^{\mathrm{High}} \bigr\}
  \end{array}
\end{cases}
\end{array}
\end{equation}
where the second condition ensures a responsive insulin suspension during hypoglycemia or when the risk of hypoglycemia rises, with $N_{\text{RoC}}^{\text{High}}$ a hyperparameter defining the negative glucose \ac{RoC} to suspend insulin delivery. The first condition enforces a conservative insulin modulation as the risk of hypoglycemia increases with aggressive insulin infusions at lower glucose levels. It is important to remark that the proposed idea of utilizing information e.g. on IoB, RoC, TDI as in~\eqref{eq:constraint2},~\eqref{eq:bounds} are not restricted to the presented constraints and can be adapted appropriately dictated by the intended therapy design, to combine actionable information in a systematic manner.

\subsubsection{The insulin bolus delivery (IBD) module} \label{subsec:SBS}
The \ac{IBD}  module is a component of the \ac{UniBE} \ac{hAID}  system designed primarily to assist the \ac{MPC} controller with meal-related insulin dosing. Its secondary role is to manage generic hyperglycemic excursions that may arise independently of meals, or in cases of substantial \ac{CHO} underestimation. The module delivers insulin boluses, both predictive and corrective, based on individual therapy parameters and BG dynamics, leveraging current and historical glucose data to anticipate and attenuate glycemic deviations. Its central task is to compute an appropriate insulin dose that compensates for unaccounted disturbances and helps restore glucose levels from $G_{\text{cur}}$ toward the target $G_{\text{tar}}$. 

To formalize the \ac{IBD}  module, we begin with its first component: the prandial insulin computation. To this aim, let us consider the standard prandial bolus $I_{\text{BP}}$ formula in (\si{U}) \cite{Schmidt2014BolusCalculators}:
\begin{equation} \label{eq:PrandialStandardBolus}
I_{\text{BP}} = \frac{\text{CHO}}{\text{CR}} + \frac{G_{\text{cur}} - G_{\text{tar}}}{CF} - \text{IoB},
\end{equation}
where \ac{CHO}  (\si{\gram}) is the carbohydrate content estimate of the announced meal and CR (\si{\gram\per U}) is the carbohydrate-to-insulin ratio, a patient-specific parameter typically adjusted by the treating physician \cite{Davidson2008BasalBolusGuidelines}. CF (\si{\milli\gram  \per\deci\liter \per U}) is a user-specific correction factor, typically adjusted by clinicians using empirical guidelines \cite{Davidson2008BasalBolusGuidelines}. Finally,  IoB (\si{U}) represents the insulin on board as the estimated amount of active insulin remaining from previous insulin injections. 

Modern \ac{CGM} devices provide an estimate of the glucose RoC, often visualized as trend arrows. These indicators can be valuable for refining the insulin dose calculation and improving dosing accuracy \cite{Scheiner2015PracticalCGM,Pettus2017rtCGMAdjustments}.
In this work, we define a trend arrow system based on the computed RoC, and use it to derive an adjusted glucose level $G_{\text{adj}}$ (\si{\milli\gram \per\deci\liter}), as proposed in \cite{Pettus2017rtCGMAdjustments} and summarized in Table \ref{table:adjustedGlucose}.
We enhance the standard prandial bolus formula of equation \eqref{eq:PrandialStandardBolus} by incorporating this trend-adjusted glucose level and an additional bolus adjustment term $I_{\text{RoC \& CF}}$ (\si{U}), yielding:
\begin{equation} \label{eq:insulinPrandialBolus}
I_{\text{BP}} = \frac{\text{CHO}}{\text{CR}} + I_{\text{RoC \& CF}} + \frac{G_{\text{adj}} - G_{\text{tar}}}{CF} - \text{IoB}.
\end{equation}
The term $I_{\text{RoC \& CF}}$ (\si{U}) is a correction component based on the current \ac{CGM} trend arrow and the patient-specific CF \cite{Aleppo2017DexcomArrows}. Its values are defined in Table \ref{table:adjustedGlucoseCF}.

In addition to prandial dosing, the \ac{IBD}  module also addresses persistent hyperglycemia. The correction bolus $I_{\text{CB}}$ (\si{U}) is computed as:

\begin{equation} \label{eq:InsulinSBS}I_{CB} = \alpha \left( \frac{G_{\text{adj}} - G_{\text{tar}}}{CF} - \text{IoB}\right).
\end{equation}
Here, $G_{\text{adj}}$ is the same adjusted glucose value used in equation \eqref{eq:insulinPrandialBolus}, as defined in Table \ref{table:adjustedGlucose}. Finally, the gain factor $\alpha$ is a hyperparameter modulated by the \ac{CGM} trend arrow. 

\begin{table}[h!]
\centering
\caption{Adjusted glucose value based on the current \ac{CGM} reading and its computed trend arrow.}
\renewcommand{\arraystretch}{1.5}

\begin{tabular}{|p{5cm}|c|}
\hline
\textbf{Computed CGM trend arrow and rate of change} & \textbf{$G_{\text{adj}}$ (\si{\milli\gram\per\deci\liter})} \\ 
\hline
Double arrow up/down:

Glucose increase or decrease of 3 or more \si{\milli\gram\per\deci\liter \per \minute} & $G_{\text{cur}} \pm 100$ \\
\hline
Arrow up/down: 

Glucose increase or decrease between 2 and 3 \si{\milli\gram\per\deci\liter \per \minute} & $G_{\text{cur}} \pm 75$ \\
\hline
Diagonal arrow: 

Glucose increase or decrease between 1 and 2 \si{\milli\gram\per\deci\liter \per \minute} & $G_{\text{cur}} \pm 50$ \\
\hline
Horizontal arrow: 

Glucose change between +1 and –1 \si{\milli\gram\per\deci\liter \per \minute} & $G_{\text{cur}}$ \\
\hline
\end{tabular}
\label{table:adjustedGlucose}
\end{table}

\begin{table}[h!]
\centering
\caption{Bolus adjustment component based on the patient-specific CF, and the current \ac{CGM} trend arrow.} 

% \renewcommand{\arraystretch}{1.5}
% \begin{tabular}{|p{3cm}|c|c|}
% \hline
% \textbf{CGM trend arrow} & \textbf{CF ((mg/dL)/U)} & \textbf{Insulin adjustment (U)} \\
% \hline

% Double arrow up/down: 

% Glucose increase/decrease of 3 or more (mg/dL)/min & $<$25 & $\pm 4.5$  \\
%                       & 25–$<$50 & $\pm 3.0$ \\
%                       & 50–$<$75 & $\pm 1.5$ \\
%                       & $\geq$75 & $\pm 1.0$ \\
% % \hline
% Arrow up/down:  

% Glucose increase/decrease between 2 and 3 (mg/dL)/min        & $<$25 & $\pm 3.0$ \\
%                       & 25–$<$50 & $\pm 2.0$ \\
%                       & 50–$<$75 & $\pm 1.0$ \\
%                       & $\geq$75 & $\pm 0.5$ \\
% \hline
% Diagonal arrow: 

% Glucose increase/decrease between 1 and 2 (mg/dL)/min        & $<$25 & $\pm 1.5$ \\
%                       & 25–$<$50 & $\pm 1.0$ \\
%                       & 50–$<$75 & $\pm 0.5$ \\
%                       & $\geq$75 & $\pm 0.25$ \\
% \hline
% Horizontal arrow: 

% Glucose increase/decrease between +1 and -1 (mg/dL)/min     & Any CF & 0 \\
% \hline
% \end{tabular}
% % 

\renewcommand{\arraystretch}{1.5}

\centering
\scriptsize
\begin{tabular}{|>{\centering\arraybackslash}p{2.4cm}|c|c|}
\hline
\textbf{CGM trend arrow} & \textbf{CF (\si{\milli\gram\per\deci\liter \per U})} & \textbf{$ I_{\text{RoC \& CF}}$ (\si{U})} \\
\hline
\multirow{4}{=}{Double arrow up/down:\\Glucose change $\geq$ 3 \si{\milli\gram\per\deci\liter \per \minute}} 
  & CF $<$25 & $\pm 4.5$ \\
  & $25 \leq \text{CF} < 50$ & $\pm 3.0$ \\
  & $50 \leq \text{CF} < 75$ & $\pm 1.5$ \\
  & CF $\geq$75 & $\pm 1.0$ \\
\hline
\multirow{4}{=}{Arrow up/down:\\Glucose change 2-3 \si{\milli\gram\per\deci\liter \per \minute} }
  & CF $<$25 & $\pm 3.0$ \\
  & $25 \leq \text{CF} < 50$ & $\pm 2.0$ \\
  & $50 \leq \text{CF} < 75$ & $\pm 1.0$ \\
  & CF $\geq$75 & $\pm 0.5$ \\
\hline
\multirow{4}{=}{Diagonal arrow:\\Glucose change 1-2 \si{\milli\gram\per\deci\liter \per \minute}} 
  & CF $<$25 & $\pm 1.5$ \\
  & $25 \leq \text{CF} < 50$ & $\pm 1.0$ \\
  & $50 \leq \text{CF} < 75$ & $\pm 0.5$ \\
  & CF $\geq$75& $\pm 0.25$ \\
\hline
\multirow{4}{=}{Horizontal arrow:\\Glucose change -1 to \\+1 \si{\milli\gram\per\deci\liter \per \minute} }
  &CF $<$25 & $0$ \\
  & $25 \leq \text{CF} < 50$ & $0$ \\
  & $50 \leq \text{CF} < 75$ & $0$ \\
  & CF $\geq$75 & $0$ \\

\hline
\end{tabular}

\label{table:adjustedGlucoseCF}
\end{table}

\subsection{In-silico datasets description}
All in-silico datasets were generated using the commercial version of the UVa/Padova \ac{T1D} simulator \cite{Man2014UVAPadovaSimulator}. The simulation framework incorporates dynamic distributions of patient-specific therapy parameters to emulate real-world variability and clinical complexity.

\subsubsection{Training and test sets}
\begin{figure}[t]
\centering %scale=0.57
\includegraphics[trim=100mm 10mm 90mm 10mm, clip, width=1\columnwidth]{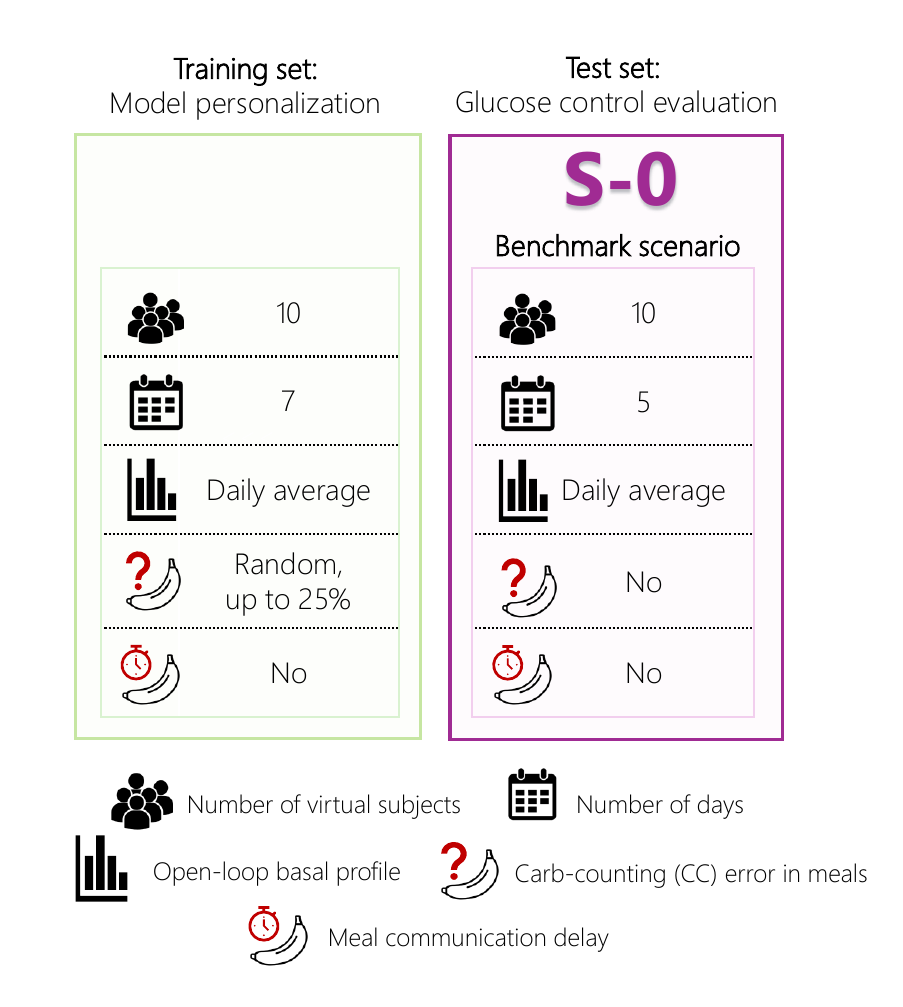} 
\caption{Overview of training and test datasets. The training set is used for model parameter personalization, while the test set is used for control performance evaluation.} 
\label{fig:blocksDatasetsTrainTest}
\end{figure}

To assess the glucose regulation capabilities of the \ac{UniBE} \ac{hAID}  system, we constructed two distinct datasets: one for training and one for testing. The training dataset was used to personalize the model for each virtual subject, while the test dataset served to evaluate glucose control performance under standard operating conditions.
Both datasets contain simulated data from 10 virtual adult \ac{T1D}  subjects, adhering to the protocol depicted in Figure \ref{fig:blocksDatasetsTrainTest}. The training dataset spans 7 consecutive days, whereas the test dataset covers a 5-day period.
\par The training dataset is designed to support model learning under nominal \ac{hAID}  operation. In this setting, the open-loop basal insulin rate is fixed at the subject’s daily average nominal value, and meals are announced promptly. To reflect real-world variability, carbohydrate estimation errors (CC-error) are introduced, modeled as a normal distribution with a standard deviation of 25\% around the true carbohydrate content.
The test dataset, by contrast, is structured to evaluate glucose control performance in an idealized scenario. Here, the basal insulin rate remains at its nominal daily value, meals are announced without delay, and carbohydrate content is reported accurately, i.e., no estimation error is introduced. This setup does not aim to replicate realistic \ac{T1D}  behavior, but rather to establish a best-case performance benchmark. It serves as a reference scenario against which all robustness and stress-test conditions are compared in subsequent analyses, as presented in Section \ref{subsec:Robustness analysis}.

\subsubsection{Robustness analysis} \label{subsec:Robustness analysis}
\begin{figure*}[t]
\centering %scale=0.57
\includegraphics[trim=0mm 10mm 0mm 10mm, clip, width=1\textwidth]{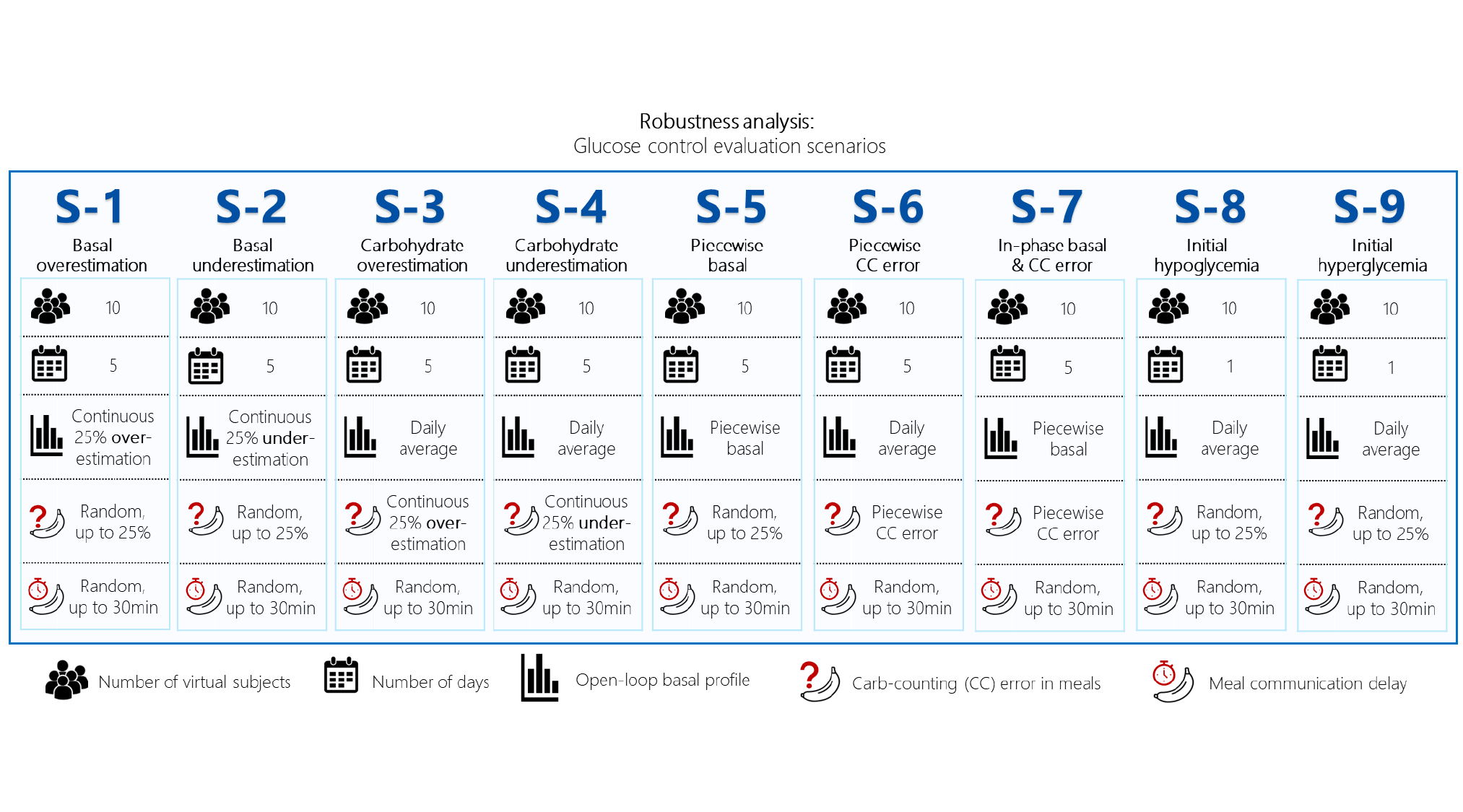} 
\captionsetup{width=\textwidth}
\caption{Overview of the robustness analysis datasets. The training set is used for model parameter personalization, while the test set is used for control performance evaluation.} 
\label{fig:blocksDatasetsRobustnessAnalysis}
\end{figure*}

To rigorously evaluate the robustness of the \ac{UniBE} hAID system, nine distinct challenge scenarios, labeled S-1 through S-9, were simulated. The nine distinct challenging scenarios are designed to test the controller’s performance under realistic yet well-defined conditions. These scenarios introduce variability and stress factors that reflect common challenges faced in everyday BG management, enabling a comprehensive assessment of the system’s adaptability and reliability, similar to the stress tests presented to the regulatory authorities for in-human clinical trial clearance. A description of the stress tests is presented in Figure \ref{fig:blocksDatasetsRobustnessAnalysis}. 

Each scenario involved ten virtual adult subjects with \ac{T1D}, simulated over varying time periods. Perturbations were introduced to evaluate the controller’s adaptability under realistic conditions. Scenario S-1 imposed a continuous 25\% overestimation of the open-loop basal insulin rate, simulating excessive basal delivery and the risk of hypoglycemia. In contrast, Scenario S-2 applied a continuous 25\% underestimation of basal insulin, challenging the system’s ability to manage hyperglycemia caused by insufficient insulin. 
Scenarios S-1 and S-2 are designed to mimic a suboptimal open-loop basal insulin profile. This reflects the common clinical situation in which basal rates are initially determined through successive approximations by clinicians. As a result, especially at the onset of therapy, the prescribed basal profile may deviate substantially from the patient’s true optimal dose \cite{nauck2021prediction}.

Scenarios S-3 and S-4 addressed carb counting (CC) errors: S-3 simulated random overestimations of meal carbohydrate content up to 25\%, while S-4 introduced random underestimations of the same magnitude, both reflecting common inaccuracies in meal reporting. These scenarios aim to reflect the fact that cc poses a significant challenge for individuals with \ac{T1D}, as meal composition, portion size, and hidden ingredients often lead to systematic or random misestimations of carbohydrate content \cite{brazeau2013carbohydrate}, which in turn can result in inaccurate insulin dosing and impaired glycemic control.

Scenario S-5 modified the open-loop basal insulin delivery with a piecewise profile: one day at the nominal daily averaged value, followed by one day with a 25\% overestimation, then a return to nominal, a subsequent day with a 25\% underestimation, and finally another nominal day. This pattern mimics scheduling inconsistencies and helps test the controller's ability to adapt to changing situations. Scenario S-6 introduced piecewise CC errors, beginning with a day of random 25\% error, followed by a day of systematic 25\% overestimation, then another day of random error, a day of systematic 25\% underestimation, and finally a day of random error, representing irregularities in meal announcements. Scenario S-7 explored in-phase carbohydrate counting errors, combining the basal profiles from S-5 with the meal announcements from S-6.
Finally, Scenarios S-8 and S-9 examined a 1-day controller performance when starting from altered glycemic states, specifically hypoglycemia and hyperglycemia, respectively.

In all the simulated scenarios, each virtual subject consumes three meals per day. Mealtimes are randomly assigned within predefined intervals: breakfast [07:00-08:30], lunch [12:00-13:30], and dinner [19:00-20:30]. Meal sizes are drawn from truncated Gaussian distributions capped to min-max values: 15-50 \si{\gram} \ac{CHO} for breakfast, 50-90 \si{\gram} \ac{CHO} for lunch, and 30-70 \si{\gram} \ac{CHO} for dinner. The eventual carb-counting error pattern follows the specific scenario under simulation. A random meal announcement delay between 0 and 30 minutes is introduced (in scenarios S-1 to S-9) to reflect realistic variability in user-system interaction.
Hypoglycemia is treated according to the \ac{ADA} 15/15 rule: 15 \si{\gram} of fast-acting carbohydrates are administered every 15 minutes when blood glucose falls below 70 \si{\milli\gram\per\deci\liter}, in accordance with current clinical consensus \cite{Battelino2023CGMConsensus,ADA2025GlycemicGoals}.

The selected scenarios are designed to challenge the controller to persistent, clinically relevant disturbances that are commonly encountered in real-world diabetes management \cite{bequette2014fault}. Unlike transient noise or isolated errors, these systematic biases, e.g., consistent misestimation of basal insulin needs or \ac{CHO} intake, require the controller to adapt over time without inducing instability or unsafe compensation. Each scenario targets a distinct regulatory challenge, testing the controller’s robustness, adaptability, and safety under prolonged deviation from nominal assumptions. 

It is important to acknowledge that the simulations employed here are not fully realistic. The virtual subjects and metabolic responses are generated within the constraints of the available simulator, which, while widely used, has intrinsic limitations in capturing the full complexity of human physiology and behavioral variability. Consequently, the results presented cannot be interpreted as predictive of in-vivo outcomes. Rather, they serve as encouraging indicators that the proposed controller is capable of handling diverse and challenging conditions in a controlled computational environment.

Finally, the benchmark scenario (S-0) plays a critical role in enabling internal performance comparisons. It provides a controlled reference condition that isolates the controller’s behavior under nominal operating assumptions. This is essential for quantifying the impact of systematic disturbances introduced in the robustness analysis (S-1 to S-9), and for assessing the controller’s ability to maintain glycemic stability under more challenging conditions.
Moreover, the inclusion of S-0 addresses a broader challenge in the evaluation of AID systems: the lack of standardization in simulation protocols across the literature. Differences in simulator configurations, virtual subject profiles, meal distributions, and disturbance models make direct comparison of results across studies inherently unreliable. Even when using the same simulator, variations in scenario design and implementation can lead to significantly different outcomes. By establishing a consistent internal benchmark, this study ensures that performance assessments are grounded in a reproducible and interpretable framework.

\subsection{Performance evaluation}
The in-silico BG control performance has been evaluated according to the latest consensus \citep{ADA2025GlycemicGoals}. Specifically, time in range (TIR, \%) quantifies the proportion of glucose values within the target interval (70-180\,mg/dL), while time above range (TAR, \%) and time below range (TBR, \%) capture the percentages of values exceeding or falling below this range, respectively. Time in tight range (TITR, \%) quantifies the proportion of glucose values within the interval (70-140\,mg/dL). With time TBR$_{<54}$ is intended the percentage of time spent with BG values less than 54 mg/dL. Similarly, TAR$_{>250}$ refers to the percentage of time spent with BG values over 250 mg/dL.  The coefficient of BG variation (CV, \%) reflects variability relative to the BG mean. Finally, hyperglycemic and hypoglycemic events, daily BG average, as well as high blood glucose index (HBGI) and low blood glucose index (LBGI) are also considered.
\section{Results} \label{sec:Results}
\subsection{UniBE hAID parameters tuning}
The glucose regulatory model utilized by the controller was personalized to each virtual subject using a seven-day training dataset, as shown in Figure \ref{fig:blocksDatasetsTrainTest}. This dataset included BG measurements, insulin administration records, and meal information subject to \ac{CHO} counting variability. Parameter selection for personalization was guided by sensitivity analysis conducted on the training data. The parameters adjusted during this process were:  $k_e$, $V_G$, $SI_1=k_{b1}/k_{a1}$,  $SI_2=k_{b2}/k_{a2}$, $\tau_D$, and $\tau_S$. 
All remaining model parameters were assigned standard population values as reported in prior studies \cite{Hovorka2004NMPC,Hovorka2002IVGTTPartitioning}, and summarized in Table \ref{tab:hovorka_params} in the Appendix. 
The glucose regulatory model utilized by the controller is personalized by minimizing the root mean square error (RMSE) between \ac{CGM}-measured and model-predicted BG values.
The personalized parameter set remained fixed throughout all subsequent test and robustness analysis scenarios.

The hyperparameters of the \ac{UniBE} control system are selected manually by trial and error to achieve the desired control performance. Specifically: $N_p = N_c =12$,  the glucose reference $r_k = 120$ \si{\milli\gram \per\deci\liter} , $Q= 10^{-6}$, $R=10^{-6}$, $N_{\text{TDI}}^{\min}=1.5$, $N_{\text{TDI}}^{\max}=3$,  and $N_{\text{RoC}}^{\text{High}}= 0.7$. The gain factor $\alpha$ of the insulin correction bolus $I_{CB}$ is fixed to $\alpha=1.5$ when the indication is a double arrow up, $\alpha=1.4$ when there is a single arrow up, $\alpha=1.3$ when the arrow is diagonal up, and $\alpha=1$ when the arrow is horizontal or pointing down. Furthermore, correction boluses are automatically administered when the current blood glucose exceeds 180 \si{\milli\gram\per\deci\liter}, at least 180 \si{\minute} have passed since the last announced meal, and at least 30 \si{\minute} have elapsed since the previous correction bolus.

\subsection{Test set: benchmark glucose control performance} \label{subsec: test set}
\begin{table}[]
\centering
\caption{Comparison of glucose and insulin-related metrics across the benchmark scenario of the test set. The results are expressed in median [interquartile ranges], mean (standard deviation).} 
\renewcommand{\arraystretch}{1.5}
\resizebox{\columnwidth}{!}{
\begin{tabular}{|l|c|} \cline{2-2}
\multicolumn{1}{c|}{} & \multicolumn{1}{c|}{\textbf{TEST SET}} \\
\hline
\textbf{Metric} & \textbf{S-0: Benchmark scenario}  \\
\hline
TIR (\%) & 97.3 [90.8 99.1], 92.0 (13.2) \\
\hline
TITR (\%) & 70.9 [67.5 78.1], 68.6 (14.6) \\
\hline
TAR (\%) & 2.4 [0.9 9.2], 7.9 (13.2)  \\
\hline
TAR$_{>250}$  (\%) & 0 [0 0], 1.7 (5.3) \\
\hline
TBR (\%) & 0 [0 0.3] 0.1 (0.2) \\
\hline
TBR$_{<54}$  (\%) & 0 [0 0], 0 (0) \\
 \hline
HBGI & 1.5 [1.1 2.0], 2.3 (2.8) \\
\hline
LBGI & 0.4 [0.2 0.6], 0.4 (0.2) \\
\hline
CV (\%) & 22.6 [20.7 23.0], 22.6 (4.5) \\
\hline
Daily BG average (\si{\milli\gram\per\deci\liter}) & 124 [122 132], 131 (18) \\
\hline
Hypoglycemia events (Number) & 0 [0 1.0], 0.4 (0.7) \\
\hline
Hyperglycemia events (Number) & 3.0 [1.0 8.0], 4.6 (4.2)  \\
\hline
People with TIR $>$70\% (\%) & 90 \\
\hline
People with TBR $<$4\% (\%) & 100  \\
\hline
% People with Time Below 54 \si{\milli\gram\per\deci\liter} $<$1\% (\%) & 100 \\
% \hline
People with TAR $<$25\% (\%) & 90 \\
\hline
People with TAR$_{>250}$  $<$5\% (\%) & 90 \\
\hhline{|==|}
TDI per kg (\si{U \per\kilo\gram}) & 0.5 [0.4 0.9], 0.6 (0.3) \\
\hline
Basal percentage of TDI (\%) & 65.1 [61.8 68.5], 64.4 (5.5) \\
\hline
Bolus percentage of TDI (\%) & 34.9 [31.5 38.2], 35.6 (5.5) \\
\hline
Correction bolus percentage of TDI (\%) & 0 [0 0], 0.3 (0.6)\\
\hline
Prandial bolus percentage of TDI (\%) & 34.7 [31.5 38.2], 35.3 (5.8) \\
\hline
\end{tabular}}
\label{table:testSetPerformance}
\end{table}
The benchmark scenario results, summarized in Table~\ref{table:testSetPerformance}, demonstrate that under idealized benchmark conditions of average daily insulin use and without errors in meal timing or CC, the \ac{UniBE} controller achieved an average \ac{TIR} of 92.0\%, with a TAR of 7.9\% and a \ac{TBR} of 0.1\%. Over the 5‑day analysis period, this corresponded to an average of 0.4 hypoglycemic events and 4.6 hyperglycemic events, indicating stable glycemic regulation with minimal excursions.

Insulin partitioning was well balanced: the controller‑modulated basal delivery accounted for 64.4\% of the TDI, while correction boluses contributed only 0.3\%. The very low reliance on correction boluses underscores the controller’s ability to proactively maintain glucose stability through basal modulation, rather than reactive dosing. This balance reflects both the efficiency of the algorithm and its capacity to sustain BG control under idealized conditions.

\subsection{Robustness analysis} \label{subsec: Robustness analysis}
The results on all robustness analysis scenarios are summarized in Table~\ref{table:robustnessAnalysisPerformance}.

Despite persistent underestimation of basal insulin (S-2) and carbs (S-4), the controller maintained stable glucose regulation, with \ac{TIR} averaging 87.7\% and 87.0\%, and \ac{TAR} averaging 12.3\% and 12.9\%, respectively. The basal percentage of TDI in S-2 on average is 53.5\% and 66.3 \% in S-4, reflecting adaptive basal modulation. In addition, the correction bolus percentage contribution increased to 0.9 and 1.0, respectively, indicating the controller's responsiveness with automated correction boluses to address basal and carbohydrates that were underestimated over the 5-day evaluation period.

On the contrary, for the challenges of persistent overestimation of basal (S-1) and carbs (S-3), the controller resulted in average \ac{TIR} of 91.8\% and 92.8\% while limiting the hypoglycemic events to 2.7 over the 5-day period. A moderate increase in LBGI is registered,  amounting to 0.7 and 0.6 on average, respectively. The basal percentage of TDI in S-1 is on average 60.8\% and 51.4\% in S-3, and the correction bolus percentage is 0.6 and 0.4\%, respectively. Compared to the underestimation scenarios, the shift in the basal contribution indicates compensatory basal modulation per needs while maintaining the primary objective of patient safety. In the persistent overestimation robustness test setting, the controller demonstrated robustness, achieving stable glucose control while maintaining \ac{TIR} and limiting hypoglycemia burden.

In scenarios S-5 to S-6, where the errors on basal and CC-counting are varied over time, the average \ac{TIR} are 91.0\% and 91.2\%, \ac{TAR} are 8.6\%  and 8.4\%, and average hypoglycemia events are 1.2 and 1.4, respectively. These indicate consistent controller performance against day-to-day varying disturbances. This is achieved with the \% of TDI being basal at 57.7\% for S-5 and 58.3\% for S-6, and correction boluses of 0.4\% and 0.2\%, respectively, reflecting slight adaptive adjustments by the controller.

Similarly, in scenario S-7 with in-phase errors in both basal and CC-counting, the controller achieved similar average performance: 89.5\% \ac{TIR}, 9.9\% \ac{TAR}, with 2.0 hypoglycemic events. The controller modulation was maintained at both basal (56.9\% of TDI) and during correction boluses, which increased to 0.6\%. This demonstrated that the controller can maintain stable glucose control under simultaneous error from basal and meals.

In scenario S-8, initial hyperglycemia and S-9 initial hypoglycemia are evaluated over 1 day of simulation, respectively. In S-8, the controller achieved an average 91.4\% \ac{TIR}, 6.5\% \ac{TAR}, and a \ac{TBR} of 2.0\%, corresponding to an average of 1.2 hypoglycemic events. These results demonstrate that the controller can restore normoglycemia while minimizing recurrent hypoglycemia. In S-9, despite the initial hyperglycemia, the average \ac{TIR} is 75.1\%, \ac{TAR} is limited to 4.6\%, and the average number of hypoglycemic events does not exceed 0.2. This demonstrates robust and safe recovery from initial hyperglycemia without inducing hypoglycemia, while reacting to the hyperglycemia corrections.

\section{Discussion} \label{sec:Discussion}
Across all evaluated scenarios, the controller demonstrated effective and stable glycemic regulation under a wide range of robustness conditions. 

In the benchmark scenario (S‑0), we want to establish a reference point for evaluating performance under induced varying degrees of disturbances (i.e., the robustness analysis).
Under the persistent bias challenges of the robustness analysis, the controller demonstrated adaptive responses. In underestimation scenarios (S‑2 and S‑4), the controller basal modulation and automatic correction boluses resulted in average \ac{TIR} above 87\%. Conversely, in overestimation scenarios (S‑1 and S‑3), stable glycemic control is achieved with a moderate increase in hypoglycemic events, highlighting the adaptability of the controller to the over-delivery of insulin due to increased basal and post-prandial boluses.
In variable and combined error scenarios (S‑5 to S‑7), where errors are considered on basal and \ac{CHO} estimation, the system demonstrated low variability in glucose control achieved, demonstrating flexibility to adapt to user-induced disturbances that varied over multiple days. This is achieved by a balanced insulin partitioning and minimal, yet adequately timed, usage of automated correction boluses, throughout dynamic error conditions. The low reliance on corrective boluses observed in the benchmark scenario underscores the controller’s ability to maintain glucose stability primarily through proactive basal modulation.
In cases of extreme initial conditions (S-8 and S-9), representing hypo- and hyper-glycemic starting BG levels, the controller is able to restore the normoglycemic levels and maintain them effectively over the whole evaluation period. In both cases, the recovery is achieved without rebound events, validating the controller's corrective response and stability under non-steady-state dynamical conditions.

Moreover, for scenarios S-0 to S-8, at least 90\% of participants satisfied all the recommended levels in terms of \ac{TIR} $>$70\%, \ac{TBR} $<$4\%, \ac{TBR}$_{<54}$ <1\%, and \ac{TAR} $<$25\%.
The glucose CV (\%) remains, on average, below the recommended value of <36\% across all scenarios \cite{ADA2025GlycemicGoals}.

Finally, it is important to underline that the present work should be regarded as a proof of concept rather than a definitive validation of clinical performance.

\section{Conclusion}
\label{sec:conclusion}
This work introduces the \ac{UniBE} controller, an advanced hybrid \ac{AID} system built upon a successive linearization \ac{MPC} framework with dynamic constraints adaptation utilizing a personalized glucose dynamical model. The \ac{UniBE} hAID was specifically designed to address a wide range of challenging scenarios in \ac{T1D} management, where variability in patient behavior and physiology often undermines the effectiveness of conventional approaches. Using the commercial version of the FDA‑accepted UVa/Padova metabolic simulator \cite{Man2014UVAPadovaSimulator}, we conducted in‑silico experiments to rigorously evaluate the controller’s adaptability under diverse perturbations.

By systematically introducing clinically relevant disturbances, including both persistent and time-varying errors in basal insulin estimation, inaccuracies in carbohydrate counting, and deviations in initial glycemic states, we aimed to approximate, though not fully replicate, the variability encountered in real-world practice. The results demonstrate that the \ac{UniBE} controller can sustain high levels of performance across these scenarios, highlighting its robustness and feasibility as a candidate for future clinical translation. 

The contribution of this study lies in establishing proof of concept: the \ac{UniBE} controller can adapt successfully to common sources of error and variability in diabetes management. These findings provide a strong foundation for further refinement of the algorithm and motivate subsequent steps toward experimental validation. Nevertheless, they should not be interpreted as conclusive evidence of clinical efficacy. Future work must extend beyond in‑silico testing to encompass preclinical and clinical studies, where the true adaptability, safety, and patient benefit of the controller can be rigorously assessed.

\appendix
\section*{List of Acronyms}
ADA – American diabetes association, AID – automated insulin delivery, BG – blood glucose, CC - carbohydrate-counting,  CF – correction factor, CGM – continuous glucose monitor, CHO – carbohydrate, CR – carbohydrate-to-insulin ratio, EASD – European association for the study of diabetes, EGP – endogenous glucose production, EKF – extended Kalman filter, FDA – food and drug administration, HBGI – high blood glucose index, IBD – insulin bolus delivery, IOB – insulin on board, LBGI – low blood glucose index, MAS – meal-autonomous system, MDI – multiple daily injections, MPC – model predictive control, PID – proportional-integral-derivative, RMSE – root mean square error, RoC – rate of change, SD – standard deviation,  T1D – type 1 diabetes, TAR – time above range,  TAR – time above range, TBR – time below range, TDI – total daily insulin, TITR – time in tight range, TITR – time in tight range, TIR – time in range, UniBE – University of Bern.
\section{Appendix}
\subsection*{The Hovorka gluco-regulatory model} \label{appendix:HovorkaModel}
The model considered in the \ac{MPC} architecture is the Hovorka glucose-insulin model \cite{Hovorka2002IVGTTPartitioning, Hovorka2004NMPC}. It can be summarized as follows: 
\begin{align*}
\frac{dQ_1(t)}{dt} &= U_G(t) - x_1(t)Q_1(t) - F_{01}^C(t) \nonumber \\
&\quad - F_R(t) + k_{12}Q_2(t) + EGP(t)
\end{align*}
\[
\frac{dQ_2(t)}{dt} = x_1(t)Q_1(t) - [x_2(t) + k_{12}]Q_2(t) 
\]
\[
\frac{dx_1(t)}{dt} = -k_{a_1}x_1(t) + k_{b_1}I(t) 
\]
\[
\frac{dx_2(t)}{dt} = -k_{a_2}x_2(t) + k_{b_2}I(t) 
\]
\[
\frac{dS_1(t)}{dt} = -\frac{S_1(t)}{\tau_S} + u(t) 
\]
\[
\frac{dS_2(t)}{dt} = \frac{S_1(t)}{\tau_S} - \frac{S_2(t)}{\tau_S} 
\]
\[
\frac{dI(t)}{dt} = -k_e I(t) + \frac{S_2(t)}{\tau_S V_I} 
\]
\[
\frac{dD_1(t)}{dt} = \frac{1000\text{AG}}{M_{wg}}d(t) -\frac{D_1(t)}{\tau_D}
\]
\[
\frac{dD_2(t)}{dt} = \frac{D_1(t)}{\tau_D} -\frac{D_2(t)}{\tau_D}
\]
\[
F^C_{01}(t) = 
\begin{cases} 
F_{01}(t), & G(t) \geq 4.5 \text{ mmol/L} \\
\frac{F_{01}(t)G(t)}{4.5}, & \text{otherwise}
\end{cases}
\]
\[
F_R(t) = 
\begin{cases} 
0.003(G(t) - 9)V_G, & G(t) \geq 9 \text{ mmol/L} \\
0, & \text{otherwise}
\end{cases}
\]
\[
EGP(t) = 
\begin{cases} 
EGP_0[1 - x_3(t)], & EGP \geq 0 \\
0, & \text{otherwise}
\end{cases}
\]
where the states $Q_1(t)$, $Q_2(t)$ (mmol) represent the glucose masses in accessible and non-accessible compartments; $x_1 (t)$, $x_2(t)$, $x_3(t)$ (\si{\per\minute}) are the effects of insulin on glucose distribution, disposal and endogenous glucose production; $S_1(t)$, $S_2 (t)$ (mU) represent absorption of injected insulin; I(t) (mU/L) represents plasma insulin concentration; $D_1(t)$, $D_2(t)$ (mmol) represents the \ac{CHO} absorption subsystem; $U_G(t)$ (mmol/min) is the gut absorption rate; $F_{01}^c(t)$ is the non-insulin dependent glucose uptake (mmol/min); $F_R(t)$ is the renal glucose excretion (mmol/min); G(t) (mmol/L) represents the plasma glucose concentration; EGP(t) represents endogenous glucose production. The inputs to the model are oral \ac{CHO} intake d(t) (g/min) and insulin injected u(t) (mU/min).

All remaining model parameters were assigned standard population values as reported in prior studies \cite{Hovorka2004NMPC,Hovorka2002IVGTTPartitioning}, and summarized in Table \ref{tab:hovorka_params}.
\begin{table}[t]
\centering
\caption{Hovorka model parameters and population values.}
\renewcommand{\arraystretch}{1.5}
\resizebox{\columnwidth}{!}{
\begin{tabular}{|l|l|l|l|}
\hline
\textbf{Parameter} & \textbf{Description} & \textbf{Unit} & \textbf{Value} \\
\hline
$k_{12}$ & Transfer rate & min$^{-1}$ &  $66\times 10^{-3}$ \\
\hline
$EGP_0/\text{BW}$ & \makecell[l]{Endogenous glucose production \\ (0 insulin concentration)} & \si{\milli mol \per\minute \per \kilo\gram
}  &  $161 \times 10^{-4}$ \\
\hline
$k_{a1}$ & Deactivation rate & min$^{-1}$ &  $6 \times 10^{-3}$ \\
\hline
$S_{I1} = k_{b1} / k_{a1}$ & \makecell[l]{ Insulin sensitivity of \\ distribution/transport } & 1/min/L/mU & $51.2 \times 10^{-4}$ \\
\hline
$k_{a2}$ & Deactivation rate & min$^{-1}$ &  $6 \times 10^{-3}$ \\
\hline
$S_{I2} = k_{b2} / k_{a2}$ & Insulin sensitivity of disposal & 1/min/L/mU & $8.2 \times 10^{-4}$ \\
\hline
$k_{a3}$ & Deactivation rate & min$^{-1}$ & 0.03 \\
\hline
$S_{I3} = k_{b3} / k_{a3}$ & Insulin sensitivity of EGP & \si{1\per\minute\per\liter\per\milli U}& $520 \times 10^{-4}$ \\
\hline
$\tau_s$ & Insulin absorption constant & min & 55 \\
\hline
$k_e$ & Insulin elimination rate & min$^{-1}$ &  $138 \times 10^{-3}$ \\
\hline
$AG$ & CHO bioavailability & -- & 0.8 \\
\hline
$M_{wg}$ & Glucose molecular weight & \si{\gram\per\mol}& 180.16 \\
\hline
$\tau_D$ & CHO absorption constant & min & 40 \\
\hline
$V_G/\text{BW}$ & Glucose distribution volume & \si{\liter\per\kilo\gram} & 0.16 \\
\hline
\end{tabular}}
\label{tab:hovorka_params}
\end{table}

\bibliographystyle{elsarticle-num}

\bibliography{myLoop}

@techreport{ADA_EASD2026,
  author = {{American Diabetes Association} and {European Association for the Study of Diabetes}},
  title = {The Management of Type 1 Diabetes in Adults: 2026 Consensus Report (Draft for Public Comment)},
  year = {2026},
  month = {},
  note = {Public comment draft. Accessed December 5, 2025},
  url = {https://professional.diabetes.org/sites/dpro/files/2025-09/2026-ADA-EASD-Management-of-Type-1-Diabetes-Consensus-Report-Draft-for-Public-Comment.pdf},
  urldate = {2025-12-05}
}

@article{AdverseOutcomesT1D2024,
title = {Acute and Chronic Adverse Outcomes of Type 1 Diabetes},
journal = {Endocrinology and Metabolism Clinics of North America},
volume = {53},
number = {1},
pages = {123-133},
year = {2024},
note = {Type 1 Diabetes},
issn = {0889-8529},
doi = {https://doi.org/10.1016/j.ecl.2023.09.004},
url = {https://www.sciencedirect.com/science/article/pii/S0889852923000634},
author = {Rachel Longendyke and Jody B. Grundman and Shideh Majidi}
}

@article{ADA2025Technology, author = {{American Diabetes Association}}, title = {7. Diabetes Technology: Standards of Care in Diabetes—2025}, journal = {Diabetes Care}, year = {2024}, volume = {48}, number = {Supplement 1}, pages = {S146--S166}, doi = {10.2337/dc25-S007} }

@article{Breton2021ControlIQ, author = {Breton, Marc D. and Kovatchev, Boris P.}, title = {One Year Real-World Use of the Control-IQ Advanced Hybrid Closed-Loop Technology}, journal = {Diabetes Technology and Therapeutics}, year = {2021}, volume = {23}, number = {9}, pages = {601--608}, doi = {10.1089/dia.2021.0097} }

@article{DaSilva2022MiniMed780G, author = {Da Silva, J. and Lepore, G. and Battelino, T. and others}, title = {Real-World Performance of the MiniMedTM 780G System: First Report of Outcomes from 4120 Users}, journal = {Diabetes Technology and Therapeutics}, year = {2022}, volume = {24}, number = {2}, pages = {113--119}, doi = {10.1089/dia.2021.0203} }

@article{Forlenza2023Omnipod5, author = {Forlenza, G. and De Salvo, D. and Aleppo, G. and others}, title = {Real-World Evidence of Omnipod 5 Automated Insulin Delivery System Use in 69,902 People with Type 1 Diabetes}, journal = {Diabetes Technology and Therapeutics}, year = {2023}, volume = {26}, number = {8}, pages = {514--525}, doi = {10.1089/dia.2023.0578} }

@article{garcia2021advanced,
  title={Advanced hybrid artificial pancreas system improves on unannounced meal response-In silico comparison to currently available system},
  author={Garcia-Tirado, Jose and Lv, Dayu and Corbett, John P and Colmegna, Patricio and Breton, Marc D},
  journal={Computer Methods and Programs in Biomedicine},
  volume={211},
  pages={106401},
  year={2021},
  publisher={Elsevier}}

@article{Hovorka2002IVGTTPartitioning,
  author = {Hovorka, R. and Shojaee-Moradie, F. and Carroll, P. V. and others},
  title = {Partitioning glucose distribution/transport, disposal, and endogenous production during IVGTT},
  journal = {Am J Physiol-Endocrinol Metab},
  year = {2002},
  volume = {282},
  number = {5},
  pages = {E992--E1007},
  doi = {10.1152/ajpendo.00304.2001}
}

@article{Hovorka2004NMPC,
  author = {Hovorka, R. and Canonico, V. and Chassin, L. J. and others},
  title = {Nonlinear model predictive control of glucose concentration in subjects with type 1 diabetes},
  journal = {Physiol Meas},
  year = {2004},
  volume = {25},
  number = {4},
  pages = {905},
  doi = {10.1088/0967-3334/25/4/010}
}

@article{Escorihuela2025ParameterRelevance,
  author = {Escorihuela-Altaba, C. and Naik, V. V. and Manzoni, E. and Garcia-Tirado, J.},
  title = {Parameters Relevance of a Glucose-Insulin Model in Type 1 Diabetes is Dependent on Meal Behavior},
  journal = {IFAC-Pap},
  year = {2025},
  volume = {59},
  number = {2},
  pages = {121--126},
  doi = {10.1016/j.ifacol.2025.06.021}
}

@book{Saltelli2008SensitivityPrimer,
  author = {Saltelli, A. and Ratto, M. and Andres, T. and others},
  title = {Global Sensitivity Analysis: The Primer},
  publisher = {John Wiley and Sons},
  year = {2008}
}

@article{Henson1998NMPCReview,
  author = {Henson, M. A.},
  title = {Nonlinear model predictive control: current status and future directions},
  journal = {Comput Chem Eng},
  year = {1998},
  volume = {23},
  number = {2},
  pages = {187--202},
  doi = {10.1016/S0098-1354(98)00260-9}
}

@techreport{Welch1995KalmanFilter,
  author = {Welch, G. and Bishop, G.},
  title = {An Introduction to the Kalman Filter},
  institution = {University of North Carolina at Chapel Hill},
  year = {1995}
}

@article{Schmidt2014BolusCalculators,
  author = {Schmidt, S. and Norgaard, K.},
  title = {Bolus Calculators},
  journal = {J Diabetes Sci Technol},
  year = {2014},
  volume = {8},
  number = {5},
  pages = {1035--1041},
  doi = {10.1177/1932296814532906}
}

@article{Davidson2008BasalBolusGuidelines,
  author = {Davidson, P. C. and Hebblewhite, H. R. and Steed, R. D. and Bode, B. W.},
  title = {Analysis of Guidelines for Basal-Bolus Insulin Dosing: Basal Insulin, Correction Factor, and Carbohydrate-To-Insulin Ratio},
  journal = {Endocr Pract},
  year = {2008},
  volume = {14},
  number = {9},
  pages = {1095--1101},
  doi = {10.4158/EP.14.9.1095}
}

@book{Scheiner2015PracticalCGM,
  author = {Scheiner, G.},
  title = {Practical CGM: Improving Patient Outcomes through Continuous Glucose Monitoring},
  publisher = {American Diabetes Association},
  year = {2015}
}

@article{Pettus2017rtCGMAdjustments,
  author = {Pettus, J. and Edelman, S. V.},
  title = {Recommendations for Using Real-Time Continuous Glucose Monitoring (rtCGM) Data for Insulin Adjustments in Type 1 Diabetes},
  journal = {J Diabetes Sci Technol},
  year = {2017},
  volume = {11},
  number = {1},
  pages = {138--147},
  doi = {10.1177/1932296816663747}
}

@article{Aleppo2017DexcomArrows,
  author = {Aleppo, G. and Laffel, L. M. and Ahmann, A. J. and others},
  title = {A Practical Approach to Using Trend Arrows on the Dexcom G5 CGM System for the Management of Adults With Diabetes},
  journal = {J Endocr Soc},
  year = {2017},
  volume = {1},
  number = {12},
  pages = {1445--1460},
  doi = {10.1210/js.2017-00388}
}

@article{Battelino2023CGMConsensus,
  author = {Battelino, T. and Alexander, C. M. and Amiel, S. A. and others},
  title = {Continuous glucose monitoring and metrics for clinical trials: an international consensus statement},
  journal = {Lancet Diabetes Endocrinol},
  year = {2023},
  volume = {11},
  number = {1},
  pages = {42--57},
  doi = {10.1016/S2213-8587(22)00319-9}
}

@article{ADA2025GlycemicGoals,
  author = {{American Diabetes Association}},
  title = {6. Glycemic Goals and Hypoglycemia: Standards of Care in Diabetes-2025},
  journal = {Diabetes Care},
  year = {2024},
  volume = {48},
  number = {Supplement 1},
  pages = {S128--S145},
  doi = {10.2337/dc25-S006}
}

@article{Man2014UVAPadovaSimulator,
  author = {Dalla Man, C. and Micheletto, F. and Lv, D. and Breton, M. and Kovatchev, B. and Cobelli, C.},
  title = {The UVA/PADOVA Type 1 Diabetes Simulator: New Features},
  journal = {Journal of Diabetes Science and Technology},
  year = {2014},
  volume = {8},
  number = {1},
  pages = {26--34},
  doi = {10.1177/1932296813514502}
}

@article{Wilinska2010Simulation,
  author    = {Malgorzata E. Wilinska and Ludovic J. Chassin and Carlo L. Acerini and Janet M. Allen and David B. Dunger and Roman Hovorka},
  title     = {Simulation Environment to Evaluate Closed-Loop Insulin Delivery Systems in Type 1 Diabetes},
  journal   = {Journal of Diabetes Science and Technology},
  volume    = {4},
  number    = {1},
  pages     = {132--144},
  year      = {2010},
  doi       = {10.1177/193229681000400117},
  publisher = {Diabetes Technology Society}
}

@article{brazeau2013carbohydrate,
  title={Carbohydrate counting accuracy and blood glucose variability in adults with type 1 diabetes},
  author={Brazeau, AS and Mircescu, H and Desjardins, K and Leroux, C and Strychar, I and Eko{\'e}, JM and Rabasa-Lhoret, R},
  journal={Diabetes research and clinical practice},
  volume={99},
  number={1},
  pages={19--23},
  year={2013},
  publisher={Elsevier}
}

@article{gregory2022global,
  title={Global incidence, prevalence, and mortality of type 1 diabetes in 2021 with projection to 2040: a modelling study},
  author={Gregory, Gabriel A and Robinson, Thomas IG and Linklater, Sarah E and Wang, Fei and Colagiuri, Stephen and de Beaufort, Carine and Donaghue, Kim C and Harding, Jessica L and Wander, Pandora L and Zhang, Xinge and others},
  journal={The lancet Diabetes \& endocrinology},
  volume={10},
  number={10},
  pages={741--760},
  year={2022},
  publisher={Elsevier}
}

@article{sherr2022automated,
  title={Automated insulin delivery: benefits, challenges, and recommendations. A Consensus Report of the Joint Diabetes Technology Working Group of the European Association for the Study of Diabetes and the American Diabetes Association},
  author={Sherr, Jennifer L and Heinemann, Lutz and Fleming, G Alexander and Bergenstal, Richard M and Bruttomesso, Daniela and Hanaire, H{\'e}l{\`e}ne and Holl, Reinhard W and Petrie, John R and Peters, Anne L and Evans, Mark},
  journal={Diabetes Care},
  volume={45},
  number={12},
  pages={3058--3074},
  year={2022},
  publisher={American Diabetes Association}
}

@article{lin2021continuous,
  title={Continuous glucose monitoring: A review of the evidence in type 1 and 2 diabetes mellitus},
  author={Lin, Rose and Brown, Fran and James, Steven and Jones, Jessica and Ekinci, Elif},
  journal={Diabetic Medicine},
  volume={38},
  number={5},
  pages={e14528},
  year={2021},
  publisher={Wiley Online Library}
}

@article{bequette2014fault,
  title={Fault detection and safety in closed-loop artificial pancreas systems},
  author={Bequette, B Wayne},
  journal={Journal of diabetes science and technology},
  volume={8},
  number={6},
  pages={1204--1214},
  year={2014},
  publisher={SAGE Publications Sage CA: Los Angeles, CA}
}

@article{ware2022closed,
  title={Closed-loop insulin delivery: update on the state of the field and emerging technologies},
  author={Ware, Julia and Hovorka, Roman},
  journal={Expert review of medical devices},
  volume={19},
  number={11},
  pages={859--875},
  year={2022},
  publisher={Taylor \& Francis}
}

@article{castle2017future,
  title={Future of automated insulin delivery systems},
  author={Castle, Jessica R and DeVries, J Hans and Kovatchev, Boris},
  journal={Diabetes technology \& therapeutics},
  volume={19},
  number={S3},
  pages={S--67},
  year={2017},
  publisher={Mary Ann Liebert, Inc. 140 Huguenot Street, 3rd Floor New Rochelle, NY 10801 USA}
}

@article{brown2019six,
  title={Six-month randomized, multicenter trial of closed-loop control in type 1 diabetes},
  author={Brown, Sue A and Kovatchev, Boris P and Raghinaru, Dan and Lum, John W and Buckingham, Bruce A and Kudva, Yogish C and Laffel, Lori M and Levy, Carol J and Pinsker, Jordan E and Wadwa, R Paul and others},
  journal={New England Journal of Medicine},
  volume={381},
  number={18},
  pages={1707--1717},
  year={2019},
  publisher={Mass Medical Soc}
}

@article{silva2022real,
  title={Real-world performance of the MiniMed™ 780G system: first report of outcomes from 4120 users},
  author={Silva, Julien Da and Lepore, Giuseppe and Battelino, Tadej and Arrieta, Arcelia and Casta{\~n}eda, Javier and Grossman, Benyamin and Shin, John and Cohen, Ohad},
  journal={Diabetes technology \& therapeutics},
  volume={24},
  number={2},
  pages={113--119},
  year={2022},
  publisher={Mary Ann Liebert, Inc., publishers 140 Huguenot Street, 3rd Floor New~…}
}

@article{leelarathna2025efficacy,
  title={Efficacy of the Omnipod 5 Automated Insulin Delivery (AID) System Compared with Multiple Daily Injections in Type 1 Diabetes: A Multinational Randomized Controlled Trial (RADIANT)},
  author={Leelarathna, L and Wilmot, E and Beltrand, J and Guerci, B and Berot, A and Hanaire, H and Bismuth, E and Gillard, P and Saade, MB and Joubert, M and others},
  journal={Diabetologie und Stoffwechsel},
  volume={20},
  number={S 01},
  pages={P14--08},
  year={2025},
  publisher={Georg Thieme Verlag KG}
}

@article{aiello2021review,
  title={Review of automated insulin delivery systems for individuals with type 1 diabetes: tailored solutions for subpopulations},
  author={Aiello, Eleonora M and Deshpande, Sunil and {\"O}zaslan, Ba{\c{s}}ak and Wolkowicz, Kelilah L and Dassau, Eyal and Pinsker, Jordan E and Doyle III, Francis J},
  journal={Current opinion in biomedical engineering},
  volume={19},
  pages={100312},
  year={2021},
  publisher={Elsevier}
}

@article{thomas2022algorithms,
  title={Algorithms for automated insulin delivery: an overview},
  author={Thomas, Andreas and Heinemann, Lutz},
  journal={Journal of diabetes science and technology},
  volume={16},
  number={5},
  pages={1228--1238},
  year={2022},
  publisher={SAGE Publications Sage CA: Los Angeles, CA}
}

@article{cinar2019automated,
  title={Automated insulin delivery algorithms},
  author={Cinar, Ali},
  journal={Diabetes Spectrum},
  volume={32},
  number={3},
  pages={209--214},
  year={2019},
  publisher={American Diabetes Association}
}

@misc{steil2008closed,
  title={Closed loop system for controlling insulin infusion},
  author={Steil, Garry M and Rebrin, Kerstin},
  year={2008},
  month=apr # "~8",
  publisher={Google Patents},
  note={US Patent 7,354,420}
}

@article{steil2013algorithms,
  title={Algorithms for a closed-loop artificial pancreas: the case for proportional-integral-derivative control},
  author={Steil, Garry M},
  journal={Journal of diabetes science and technology},
  volume={7},
  number={6},
  pages={1621--1631},
  year={2013},
  publisher={SAGE Publications Sage CA: Los Angeles, CA}
}

@article{nimri2014night,
  title={Night glucose control with MD-Logic artificial pancreas in home setting: a single blind, randomized crossover trial—interim analysis},
  author={Nimri, Revital and Muller, Ido and Atlas, Eran and Miller, Shahar and Kordonouri, Olga and Bratina, Natasa and Tsioli, Christiana and Stefanija, Magdalena A and Danne, Thomas and Battelino, Tadej and others},
  journal={Pediatric diabetes},
  volume={15},
  number={2},
  pages={91--99},
  year={2014},
  publisher={Wiley Online Library}
}

@article{Breton2012AP,
  author    = {Breton, M. and Farret, A. and Bruttomesso, D. and et al.},
  title     = {Fully integrated artificial pancreas in type 1 diabetes: modular closed-loop glucose control maintains near normoglycemia},
  journal   = {Diabetes},
  year      = {2012},
  volume    = {61},
  number    = {9},
  pages     = {2230--2237},
  doi       = {10.2337/db11-1406}
}

@article{Turksoy2018MultimoduleAP,
  author    = {Turksoy, K. and Littlejohn, E. and Cinar, A.},
  title     = {Multimodule multivariable artificial pancreas for patients with type 1 diabetes},
  journal   = {IEEE Control Systems Magazine},
  year      = {2018},
  volume    = {38},
  number    = {1},
  pages     = {105--124},
  doi       = {10.1109/MCS.2017.2741479}
}

@article{Pinster2016MPCvsPID,
  author    = {Pinsker, J. E. and Lee, J. B. and Dassau, E. and et al.},
  title     = {Randomized crossover comparison of personalized MPC and PID control algorithms for the artificial pancreas},
  journal   = {Diabetes Care},
  year      = {2016},
  volume    = {39},
  number    = {7},
  pages     = {1135--1142},
  doi       = {10.2337/dc15-2817}
}

@article{boughton2024role,
  title={The role of automated insulin delivery technology in diabetes},
  author={Boughton, Charlotte K and Hovorka, Roman},
  journal={Diabetologia},
  volume={67},
  number={10},
  pages={2034--2044},
  year={2024},
  publisher={Springer}
}

@article{nauck2021prediction,
  title={Prediction of individual basal rate profiles from patient characteristics in type 1 diabetes on insulin pump therapy},
  author={Nauck, Michael A and Kahle-Stephan, Melanie and Lindmeyer, Anna M and Wenzel, Sina and Meier, Juris J},
  journal={Journal of Diabetes Science and Technology},
  volume={15},
  number={6},
  pages={1273--1281},
  year={2021},
  publisher={SAGE Publications Sage CA: Los Angeles, CA}
}

\begin{landscape}

\begin{table}[p]
  \centering 
  \caption{Comparison of glucose and insulin-related metrics across the scenarios of the robustness analysis. The results are expressed in median [interquartile ranges], mean (SD).}
  \includegraphics[trim=5mm 0mm 5mm 0mm, clip, width=1.2\textwidth]{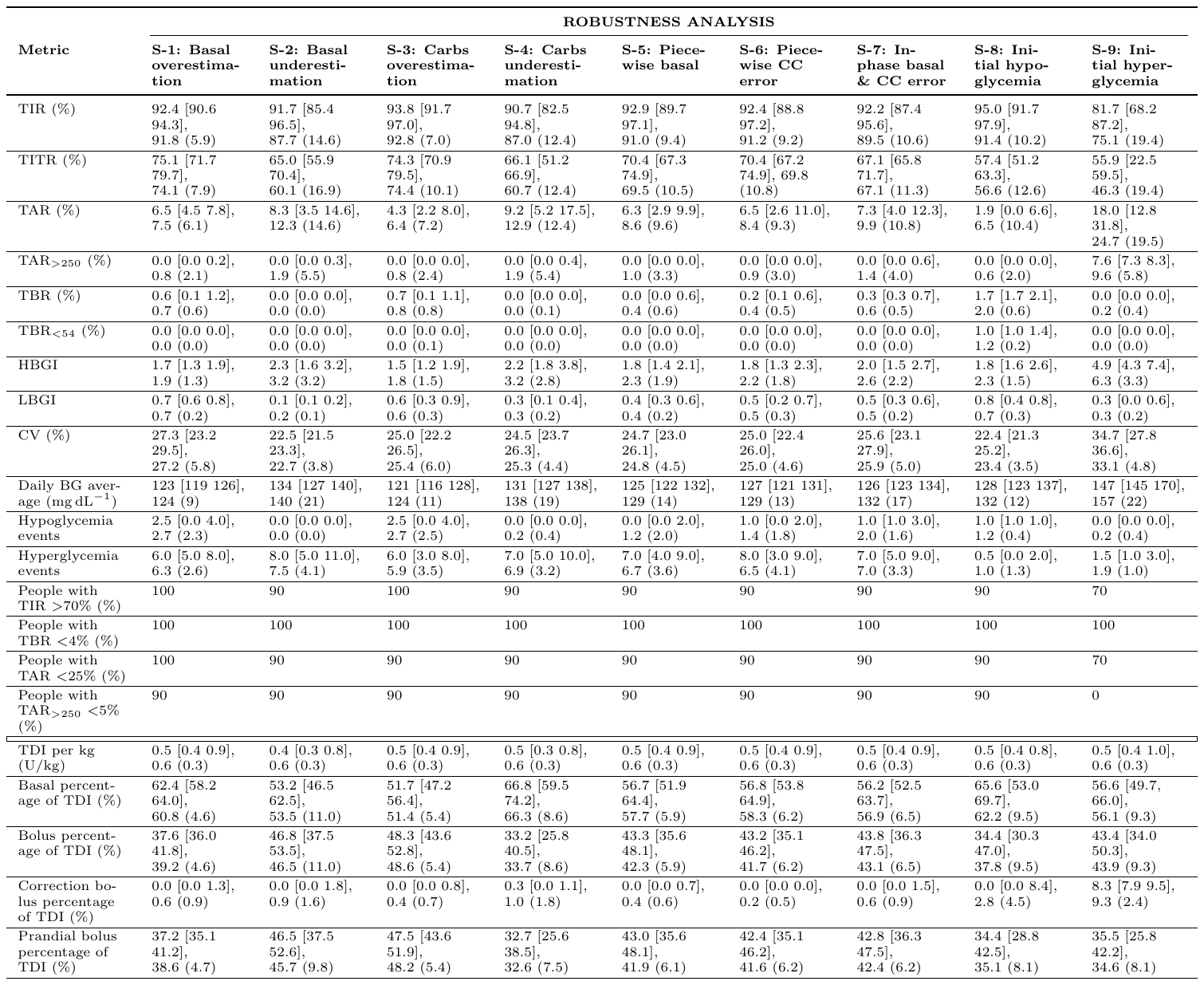}

  \label{table:robustnessAnalysisPerformance}
\end{table}
\end{landscape}

\end{document}